\documentclass[11pt]{article}
\usepackage{amsmath,amsthm,amssymb,graphicx}
\usepackage{algorithm,algorithmicx,algpseudocode}
\usepackage[a4paper,left=3cm,right=3cm,top=2cm,bottom=2cm]{geometry}
\usepackage{float}

\usepackage[colorlinks=true,urlcolor=blue, linkcolor =blue,citecolor=blue]{hyperref}

\theoremstyle{plain}

\newtheorem{remark}{Remark}

\allowdisplaybreaks

\makeatletter
\renewcommand{\@biblabel}[1]{}
\renewenvironment{thebibliography}[1]
     {\section*{\refname}%
      \@mkboth{\MakeUppercase\refname}{\MakeUppercase\refname}%
      \list{}%
           {\labelwidth=0pt
            \labelsep=0pt
            \leftmargin1em
            \itemindent=-1em
            \advance\leftmargin\labelsep
            \@openbib@code
            }%
      \sloppy
      \clubpenalty4000
      \@clubpenalty \clubpenalty
      \widowpenalty1000%
      \sfcode`\.\@m}
\makeatother
\usepackage{breakcites}

\allowdisplaybreaks

\begin{document}

\title{Efficient adaptive MCMC implementation for Pseudo-Bayesian quantum tomography}
\author{The Tien Mai}
\date{
\small
Oslo Centre for Biostatistics and Epidemiology, 
\\
Department of Biostatistics, 
\\
University of Oslo, Norway.
\\
Email: t.t.mai@medisin.uio.no
}
\maketitle

\begin{abstract}
We revisit the Pseudo-Bayesian approach to the problem of estimating density matrix in quantum state tomography in this paper.  Pseudo-Bayesian inference has been shown to offer a powerful paradign for quantum tomography with attractive theoretical and empirical results.  However,  the computation of (Pseudo-)Bayesian estimators,  due to sampling from complex and high-dimensional distribution, pose significant challenges that hampers their usages in practical settings.  To overcome this problem, we present an efficient adaptive MCMC sampling method for the Pseudo-Bayesian estimator.  We show in simulations that our approach is substantially faster than the previous implementation  by at least two orders of magnitude which is significant for practical quantum tomography.
\end{abstract}


\section{Introduction}
Quantum state tomography is a fundamental important step in quantum information processing \cite{nielsen2010quantum,MR2181445bookParis}. In general, it aims at finding the underlying density matrix which describing the given state of a physical quantum system.  This task is done by utilizing the results of measurements performed on repeated state preparations \cite{nielsen2010quantum}.

Bayesian methods have been recognized as a powerful paradigm for quantum state tomography \cite{blume2010optimal}, that deal with uncertainty in meaningful and informative ways and are the most accurate approach with respect to the expected error (operational divergence) even with finite samples.  Several studies have been conducted: for example,  the papers~\cite{buvzek1998reconst,Baier_comparison} performed numerical comparisons between Bayesian estimations with other methods on simulated data; algorithms for computing Bayesian estimators have been discussed in~\cite{kravtsov2013experimental,
ferrie2014quantum,ferrie2015have,schmied2014quantum,lukens2020practical}.

Pseudo-Bayesian method for quantum tomography, introduced in \cite{mai2017pseudo}, propose a novel approach for this problem with several attractive features.  Importantly,  a novel prior distribution for quantum density matrix is introduced based on spectral decomposition parameterization (inspired by the priors used for low-rank
matrix estimation, e.g.~\cite{mai2015,cottet20161}).  This prior can be easily used in any dimension and is found to be significantly more efficient to sample from and evaluate than the Cholesky approach in  \cite{struchalin2016experimental,zyczkowski2011generating,seah2015monte}, see \cite{lukens2020practical} for more detals.  By replacing the likelihood with a loss function between a proposed density matrix and experimental data,  the paper \cite{mai2017pseudo} presents two different estimators: the prob-estimator and the dens-estimator. 

However,  the reference \cite{mai2017pseudo} propose simply to approximate these two Pseudo-Bayesian estimators  by naive Metropolis-Hastings algorithms which is very slow for high-dimensional systems. Recently,  a faster and more efficient sampling method has been proposed for the dens-estimator, see \cite{lukens2020practical}. However, we would like to note that the prob-estimator is shown in \cite{mai2017pseudo} to reach the best known up-to-date rate of convergence \cite{butucea2015spectral} while the theoretical guarantee for the dens-estimator is far less satisfactory. Moreover, it is also shown in simulations that the prob-estimator yields better results compare to the den-estimator.

In this paper, we present a novel efficient adaptive Metropolis-Hastings implementation for the prob-estimator.  This adaptive implementation base on considering the whole density matrix as a parameter need to sample at a time. Moreover,  an adaptive proposal is explored based on the  'preconditioned Crank-Nicolson' \cite{cotter2013mcmc} sampling procedure that can elimiate the 'curse of dimensionality', which is the case for quantum state tomography where the dimension increases exponentially.  We show in the simulations that our implementation is significantly faster than the algorithm in \cite{mai2017pseudo}.

The rest of the paper is organized as follow.  In Section  \ref{sc_background},  we provide the necessary background and the statistical model for the problem  of quantum state tomography. In Section \ref{sc_pQST},  we recall the Pseudo-Bayesian approach and the prior distribution.  Section \ref{sc_implement} presents our novel adaptive MCMC implementation for the Pseudo-Bayesian estimator.  Simulations studies are presented in Section \ref{sc_num}.  Conclusions are given in Section \ref{sc_conclusion}.

\section{Background}
\label{sc_background}

\subsection{The quantum state tomography problem}
Hereafter, we only provide the necessary background on quantum state tomography required for the paper.  We would like to remind that a very nice introduction to this problem,  from a statistic perspective,  can be found in~\cite{artiles}.  Here, we have opted for the notations used in reference \cite{mai2017pseudo}.

Mathematically speaking, a two-level quantum system of $n$-qubits is characterized by a $ 2^{n}\times 2^{n} $ density matrix $\rho $ whose its entries is complex, i.e. $ \rho \in \mathbb{C}^{2^{n}\times 2^{n} } $. For the sake of simplicity,  put $d=2^n$, so $\rho$ is a $d\times d$ matrix.
This density matrix must satisfy that it is
\begin{itemize}
\item[•]  Hermitian: $\rho^\dagger=\rho$ (i.e. self-adjoint),
\item[•] positive semi-definite: $\rho \succcurlyeq 0$,
\item[•]  normalized: ${\rm Trace}(\rho)=1$.
\end{itemize}
In addition, physicists are especially interested in pure
states and that a pure state $ \rho $ can be defined by
${\rm rank}(\rho)=1$. In practice,  it often makes sense to assume that the rank of $\rho$ is small \cite{gross2010quantum,gross2011recovering,butucea2015spectral}.

The goal of quantum tomography is to estimate the underlying density matrix $\rho$ using measurement outcomes of many independent and identically systems prepared in the state $\rho$ by the same experimental devices.

For a qubit,  it is a standard procedure to measure one of the three Pauli observables $\sigma_x, \, \sigma_y, \, \sigma_z$. The outcome for each will be $ 1 $ or $ -1 $, randomly (the corresponding probability is given in~\eqref{born_rule_prob_fomala} below). As a consequence,  with a $ n $-qubits system,  there are $3^n$ possible experimental observables. The set of all possible performed observables is
\begin{align*}
\{\sigma_{\mathbf{a}} = \sigma_{{a}_1} \otimes
\ldots \otimes \sigma_{{a}_n}; \,  \mathbf{a} =
(a_1,\ldots,a_n) \in \mathcal{E}^n := \{x,y,z\}^{n}\},
\end{align*}
where vector $\mathbf{a} $ identifies the experiment.
The outcome for each fixed observable
setting will be a random vector
$ \mathbf{s} = (s_1, \ldots , s_n)   \in\{-1,1\}^{n} $,
thus there are  $ 2^n $ outcomes in total.

Denote $R^{\mathbf{a}}$ a random vector that is the outcome of an experiment indexed by $\mathbf{a}$. From the Born's rule \cite{nielsen2010quantum},  its probability distribution is given by
\begin{equation}
\label{born_rule_prob_fomala}
\forall \mathbf{s} \in\{-1,1\}^{n},
p_{\mathbf{a},\mathbf{s}} := \mathbb{P}
 (R^\mathbf{a}= \mathbf{s})
  = {\rm Trace} \left(\rho \cdot
 P_{\mathbf{s}}^{\mathbf{a}} \right),
\end{equation}
where  $ P_{\mathbf{s}}^{\mathbf{a}} :=
P_{s_1}^{a_{1}}\otimes \dots \otimes P_{s_n}^{a_n}$ and
$P_{s_i}^{a_i}$ is the orthogonal
projection associated to the eigenvalues $ s_i\in \{ \pm 1 \} $ in
the diagonalization of $ \sigma_{a_i ; , a_i\in \{x,y,z\} } $ --
that is  $ \sigma_{a_i} =  1P^{a_i}_{+1} -1P^{a_i}_{-1} $.

Statistically,  for {\it each} experiment  $ \mathbf{a}\in\mathcal{E}^n$, the experimenter
repeats $ m $ times the experiment corresponding to $\mathbf{a}$
and thus collects $m$ independent random copies of $R^\mathbf{a}$,
say $R^\mathbf{a}_1,\dots,R^\mathbf{a}_m$. As there are $3^n$
possible experiment settings $ \mathbf{a}$, we define the
\textbf{quantum sample} size as $ N:=m\cdot 3^n $. We will refer to
$(R^\mathbf{a}_i)_{i\in\{1,\dots,m\},\mathbf{a}\in\mathcal{E}^n}$
as $\mathcal{D}$ (for data). Therefore, quantum state tomography is aiming at estimating the density matrix $ \rho $ based on the data $\mathcal{D}$.

\subsection{Popular estimation methods}
Here, we briefly recall three classical major approaches have been adopted to estimate $ \rho $ which are: linear inversion, maximum likelihood and Bayesian inference.

\subsubsection*{Linear inversion}
The first and simplest method considered in quantum information processing is the 'tomographic' method also known as linear/direct inversion~\cite{vogel1989determination,vrehavcek2010operational}. It is actually the analogous of the least-square estimator in the quantum setting. This method relies on the fact that measurement outcome probabilities are linear functions of the density matrix. 

More specifically, let us consider the empirical frequencies
$$ \hat{p}_{\mathbf{a},\mathbf{s}}
= \frac{1}{m}\sum_{i=1}^m \mathbf{1}_{\{R_i^\mathbf{a}=\mathbf{s}\}}.
$$
It is noted that $ \hat{p}_{\mathbf{a},\mathbf{s}}$ is an unbiased estimator of the underlying probability $ p_{\mathbf{a},\mathbf{s}} $ in \eqref{born_rule_prob_fomala}.  Therefore, the inversion method is based on solving the linear
system of equations
\begin{equation}
 \label{rhohat}
\left\{ \begin{array}{l}
\hat p_{\mathbf{a},\mathbf{s}} = {\rm Trace}
\left(\hat {\rho} \cdot
 P_{\mathbf{s}}^{\mathbf{a}} \right),
 \\
 \mathbf{a}\in\mathcal{E}^n,\quad
 \mathbf{s} \in\{-1,1\}^{n}.
 \end{array}
 \right.
\end{equation}
As mentioned above, the computation of $\hat{\rho}$ is quite clear and explicit formulas are classical that can be found for example in e.g.~\cite{alquier2013rank}.  While straightforward and providing unbiased estimate~\cite{PhysRevLett.114.080403}, it tends to generate a non-physical density matrix as an output~\cite{shang2014quantum}: positive semi-definiteness cannot easily be satisfied and enforced.

\subsubsection*{Maximum likelihood}
A popular approach in QST in recent years is the maximum likelihood  estimation (MLE). MLE aims at finding the density matrix which is most likely to have produced the observed data $\mathcal{D}$: 
\begin{equation*}
\rho_{MLE} = \arg \max L(\rho;\mathcal{D}) 
\end{equation*}
where $ L(\rho;\mathcal{D})  $ is likelihood, the probability of observing the outcomes given state $ \rho $, as defined by some model \cite{hradil20043,james2001mea,gonccalves2018bayesian}.  However, it has some critical problems, detailed in~\cite{blume2010optimal}, including a huge computational cost.  Moreover, it is a point estimate which does not account the level of uncertainty in the result.

Furthermore, these two methods (Linear inversion and MLE) can not take advantage of a prior knowledge where a system is in a state $\rho$ for which some additional information is available.  More particularly, it is noted that physicists usually focus on so-called pure states, for which ${\rm rank}(\rho)=1$.

\subsubsection*{Bayesian inference}
Starting receiving attention in recent years, Bayesian QST had been shown as a promissing method in this problem \cite{blume2010optimal,buvzek1998reconst,Baier_comparison,lukens2020practical}. Through Bayes' theorem, experimental uncertainty is explicitly accounted in Bayesian estimation. More specifically, suppose a density matrix $ \rho $ is parameterized by $ \rho(x) $ for some $ x $, Bayesian inference is carried out via the posterior distribution
\begin{align*}
\pi( \rho(x) | \mathcal{D} ) \propto   L(\rho(x);\mathcal{D}) \pi(x),
\end{align*} 
where $  L(\rho(x);\mathcal{D}) $ is the likelihood (as in MLE) and $  \pi(x) $ is the prior distribution. Using the posterior distribution $ \pi(\rho( x) | \mathcal{D} )  $, the expectation value of any function of $ \rho $ can be inferred, e.g. the Bayesian mean estimator as $ \int \rho(x) \pi(\rho( x) | \mathcal{D}) dx  $.

Although recognized as a powerful approach,  the numerical challenge of sampling from a high-dimensional probabilty distribution prevents widespread use of Bayesian methods in the physical problem.

\subsubsection*{Other approaches}
Several other methods have also recently introduced and studied.
The reference \cite{cai2015Pauli} proposed a method based on the expansion of the density matrix $ \rho $ in the Pauli basic.  Some rank-penalized approaches were studied in ~\cite{guctua2012rank,alquier2013rank}. A thresholding method is introduced in \cite{butucea2015spectral}.

\section{Pseudo-Bayesian quantum state tomography}
\label{sc_pQST}

\subsection{Pseudo-Bayesian estimation}

Let us consider the pseudo-posterior, studied in \cite{mai2017pseudo},  defined by
\begin{align*}
\tilde{\pi}_{\lambda}({\rm d}\nu)
\propto \exp\left[-\lambda \ell(\nu,\mathcal{D}) \right]
 \pi({\rm d}\nu),
\end{align*}
where $\exp\left[-\lambda \ell(\nu,\mathcal{D}) \right]$ is the pseudo-likelihood that acting the role of the empirical evidence to give more weight to the density $\nu$  when it fits the data well;  $\pi({\rm d}\nu)$ is the prior given in Section \ref{sc_prior}; and $\lambda>0$ is a tunning parameter that balances between evidence from the data and prior information.

Taking 
$$ 
\ell(\nu,\mathcal{D})
 := 
\ell^{prob}(\nu,\mathcal{D})
 = 
 \sum_{\mathbf{a}\in\mathcal{E}^n}
\sum_{\mathbf{s}\in \mathcal{R}^n}
\left[{\rm Tr}(\nu P_\mathbf{s}^\mathbf{a})
-
\hat{p}_{\mathbf{a},\mathbf{s}}   \right]^2,
$$
the "prob-estimator" in \cite{mai2017pseudo} is defined as the mean estimator of the pseudo-posterior :
\begin{align}
\tilde{\rho}^{prob}_{\lambda}
= 
\int \nu \exp\left[-\lambda \ell^{prob}(\nu,\mathcal{D})
 \right] \pi({\rm d}\nu),
\label{prob_estimator}
\end{align}
this estimator also refered to, in statistical machine learning, as Gibbs estimator, PAC-Bayesian estimator or EWA, for exponentially weighted aggregate~\cite{catonibook,dalalyan2008aggregation}.

For the sake of simplicity, we use the shortened notation
$
p_\nu := [{\rm Tr}(\nu P_\mathbf{s}^
\mathbf{a})]_{\mathbf{a},\mathbf{s}}
$ 
and
$ \hat{p} := [\hat{p}_{\mathbf{a},
\mathbf{s}}]_{\mathbf{a},\mathbf{s}} $
 then
$$ \ell^{prob}(\nu,\mathcal{D}) = \| p_\nu -
\hat{p}  \|^2_F $$
($ \|\cdot\|_F $ is the Frobenius norm). Clearly, we can see that this distance measures the difference between the probabilities and the empirical frequencies in the sample.  We remind that the matrix $  [\hat{p}_{\mathbf{a}, \mathbf{s}}]_{\mathbf{a},\mathbf{s}} $ is of dimension $ 3^n \times 2^n $.

\begin{remark}
This kind of pseudo-posterior is an increasingly popular approach in Bayesian statistics and machine learning, see for example \cite{bissiri2013general,mai2021numerical,grunwald2017inconsistency,mai2021bayesian,catonibook,alquier2015properties,mai2021efficient,begin2016pac},  for models with intractable likelihood or for misspecification models.   
\end{remark}

\subsection{Prior distribution for quantum density matrix}
\label{sc_prior}
The pior distribution employed in \cite{mai2017pseudo} is as follow: the $ d\times d $ density matrix $ \rho $ can be parameterized by $ d $ non-negative real numbers $y_i $ and $ d $ complex column vectors of length $ d $, $z_i $.  Put $x = \left\lbrace y_1, \ldots , y_d,  z_1 , \ldots ,  z_d  \right\rbrace $, then 
the density matrix is
\begin{align}
\rho(x) = \sum_{i=1}^d \dfrac{y_i}{ \sum_\ell y_\ell } \dfrac{z_i z_i^\dagger}{\|z_i\|^2}
\end{align}
the prior distribution for $ x $ as
\begin{align}
\pi (x) \propto \prod_{i =1}^d y_i^{\alpha - 1} e^{-y_i} e^{-\frac{1}{2}z_i^\dagger z_i}
\label{prior_formula}
\end{align}
where the weights are being treated as Gamma-distributed random variables $ Y_i \overset{i.i.d.}{\sim} \Gamma(\alpha,1) $, and the vectors $ z_i $ are standard-normal complex Gaussian distributed $ Z_i \overset{i.i.d.}{\sim} \mathcal{CN} (0, I_d) $.

The tunning parameter $ \alpha $ in \eqref{prior_formula} allows the user to favor low-rank or high-rank density matrices which are corresponding to pure or mixed states, respectively.  More particularly, the normalized random variables $ Y_i /(\sum Y_j) $ with $ Y_i \overset{i.i.d.}{\sim} \Gamma(\alpha,1) $ follows a Dirichlet distribution $ {\rm Dir} (\alpha) $ which ensures both normalization and non-negativity.  An $ \alpha<1 $ promotes sparse draws and thus purer states, while $ \alpha=1 $ returns a fully uniform prior on all physically realizable states.

\begin{remark}
It is noted that this parameterization satisfies all physicallity conditions for the density matrix.  The details can be found in \cite{mai2017pseudo}. Moreover, this parameterization have been shown to be significantly more efficient to sample from and to evaluate than the Cholesky approach in references \cite{struchalin2016experimental,zyczkowski2011generating,seah2015monte}, see \cite{lukens2020practical} for details.
\end{remark}

\begin{remark}
The theoretical guarantees for the "prob-estimator" in \eqref{prob_estimator} are validated only for $0< \alpha \leq 1$. More specifically,  the prob-estimator satisfies (up to a multiplicative logarithmic factor) that $ \| \tilde{\rho}^{prob}_{\lambda^*} - \rho^0 \|_F^2  \leq c 3^n {\rm rank}(\rho^0)/N $ which is the best known up-to-date rate in the problem of quantun state estimation \cite{butucea2015spectral}, where $c$ is a numerical constant and $\lambda^* = m/2$.
\end{remark}

\section{A novel efficient adaptive MCMC Implementation}
\label{sc_implement}
Appropriately, the prob-estimator requires an evaluation of the integral \eqref{prob_estimator} which is numerically challenging due to its sophisticated features and high dimentionality.  A first attempt has been done in \cite{mai2017pseudo} is to use a naive Metropolis-Hastings (MH) algorithm where the authors iterate between a random walk MH for $Y_i$ and an independent MH for $z_i$.  Typically,  the approach is designed to obtain $T $ samples $ x^{(1)}, \ldots,   x^{(T)} $ as a consequence the integral \eqref{prob_estimator} can be approximated as
$$
\hat{\rho}^{{\rm MH}} \approx \frac{1}{T}\sum_{t=1}^T\rho (x^{(t)}).
$$
However, as also noted in the reference \cite{mai2017pseudo}, their proposed algorithm can run into slow convergence and can be arbitrarily slow as the system dimensionality increases. For sake of self-containedness,  the implementation of reference \cite{mai2017pseudo} is given in the \hyperref[appendix]{Appendix}.

Borrowing motivation from the recent work \cite{lukens2020practical} that proposes an efficient 'preconditioned Crank-Nicolson' \cite{cotter2013mcmc} sampling procedure for Bayesian quantum state estimation (which improve the computation of "dens-estimator" in \cite{mai2017pseudo} only),  we introduce an efficient adaptive Metropolis-Hastings implementation for the prob-estimator in \cite{mai2017pseudo}.  We remind that the prob-estimator shows better performance than the dens-estimator both in theory and simulations. 

Specifically, we propose to use a modification of random-walk MH by scaling the previous step before adding a random move and generating the proposal $ z' $.  Following \cite{cotter2013mcmc} who introduced an efficient MCMC approach elimiating the 'curse of dimensionality', termed as 'preconditioned Crank-Nicolson',  we use the proposal for $z_j$ as $ z_{j}^{\prime}  =  \sqrt{1-\beta_z^2} z_{j}^{(k)} + \beta_z\boldsymbol{\xi}_j ,  \boldsymbol{\xi}_j \stackrel{\textrm{i.i.d.}}{\sim} \mathcal{CN}(0,I_d) $ where $ \beta_z \in (0,1) $ is a tunning parameter.  The proposal is a scaled,  by the factor $  \sqrt{1-\beta_z^2}  $,  random walk that results in a slightly simpler acceptance probability. Unlike the independent proposal in \cite{mai2017pseudo} (with $ \beta_z=1 $) where the acceptance probability can vary substantially,  this kind of adaptive proposal allows one to control the acceptance rate efficiently. 

The acceptance ratio $ \min\{1,  A(x' | x^{(k)})  \} $ are followed from the standard form for MH \cite{robert2013monte}.  Let $ p(x' | x^{(k)} ) $ denote the proposal density, we have
\begin{align*}
A(x' | x^{(k)}) = \dfrac{ \tilde{\pi}(\rho(x')) }{ \tilde{\pi}(\rho(x^{(k)})) }
 \dfrac{ p(x^{(k)} | x') }{ p(x' | x^{(k)} ) }.
\end{align*}  
The details of the adaptive MH is given in Algorithm \ref{al_randomMH}.

\begin{algorithm}
\caption{Adaptive MH for Pseudo-Bayesian Quantum state estimation}
\label{al_randomMH}
\begin{algorithmic}
\State  \textbf{Input}:  The tunning parameters $\beta_y,\beta_z \in(0,1)$. 
\State \textbf{Parameters}: Positive real numbers $\alpha \in (0,1],T$ .
\State  \textbf{Onput}: The density matrix $\hat{\rho} $.
\State \textbf{Initialize}: $x^{(0)}$ drawn from the prior \eqref{prior_formula} ,  $ \hat{\rho} = \bold{0}_{d\times d} $.
\For{$k = 1$ to $T$} 
\State Simulate a new point $x^{\prime}=\{y_1^\prime,...,y_d^\prime;z_1^\prime,...,z_d^\prime\}$, according to
$$
\begin{aligned}
y_j^\prime & =  y_j^{(j)} e^{\beta_y\eta_j} ,
\\
z_{j}^{\prime} & =  \sqrt{1-\beta_z^2} z_{j}^{(k)} + \beta_z\boldsymbol{\xi}_j ,
\end{aligned}
\quad j \in \{1,...,d \} ,
$$
where $\eta_j\stackrel{\textrm{i.i.d.}}{\sim} Uinform(-0.5,0.5)  $ and independently $\boldsymbol{\xi}_j \stackrel{\textrm{i.i.d.}}{\sim} \mathcal{CN}(0,I_d) $.
\State Set $ x^{(k+1)}= x^\prime$ with probability $ \min \left\lbrace 1, A( x^\prime,x^{(k)}) \right\rbrace $, where
\begin{equation*}
\log A( x^\prime,x^{(k)}) 
=
 \log L_D(x^\prime) - \log L_D(x^{(k)}) + \sum_{j=1}^d \left[\alpha\log y_j^\prime - y_j^\prime -\alpha\log y_j^{(k)} + y_j^{(k)} \right] .
\end{equation*}
Otherwise set $ x^{(k+1)} =  x^{(k)}$. 
\State  $\hat{\rho} \gets  \hat{\rho} +  \rho(x^{(k)})/T $
\EndFor
\end{algorithmic}
\end{algorithm}

\section{Numerical studies}
\label{sc_num}

\subsection*{Simulations setups and details}
To access the performance of our new proposed algorithm, a series of experiments were conducted with simulated tomographic data. More particularly, we consider the following settiing for choosing the true density matrix, with $ n=2,3,4, $ ($d=4,8,16 $):
\begin{itemize}
\item Setting 1: we consider the ideal entangled state which characterized by a rank-$2 $ density matrix that
$$
 \rho_{rank-2} = \frac{1}{2}\psi_1 \psi_1^{\dagger}
+  \frac{1}{2}\psi_2 \psi_2^{\dagger}
$$ 
with $ \psi_1 = u /\| u\| $ and $ u = (u_1, \ldots, u_{d/2}, 0,\ldots,0 ),  u_1 = \ldots = u_{d/2} =1 $;  $ \psi_2 = v /\| v\| $ and $ v = (0,\ldots,0, v_{d/2 +1},  \ldots , v_d ),  v_{d/2} = \ldots = v_d = 1 $.
\item Setting 2: a maximal mixed state (rank-$d$) that is
$$
 \rho_{mixed} = \sum_{i=1}^d \frac{1}{d}\psi_i \psi_i^{\dagger},
$$
with $ \psi_i $ are normalized vectors and independently simulated from $ \mathcal{CN}(0,I_d) $.
\end{itemize}

The experiments are done following Section \ref{sc_background} for $ m=1000 $. The prob-estimator is employed with $ \lambda = m/2 $ and a prior with $ \alpha =1 $.  We compare our adaptive MH implementation, denoted by "a-MH",  against the (random-walk) in \cite{mai2017pseudo}, denoted by "r-MH".  We run 100 independent samplers for each experiment, and compute the mean of the square error (MSE),
 $$
{\rm MSE}:= \|\hat{\rho}-\rho\|_F^2 / d^2
 $$ 
 for each method, together with  their standard deviations. We also measure the mean absolute error of eigen values (MAEE) by
  $$
{\rm MAEE}:= \frac{1}{d} \sum_{i=1}^d | \lambda_i (\hat{\rho}) - \lambda_i (\rho) |,
 $$ 
 where $ \lambda_i (A) $ are the eigen values of the matrix $A $.

\subsection*{Significantly speeding up}

\begin{figure}[!ht]
\centering
\includegraphics[scale=.7]{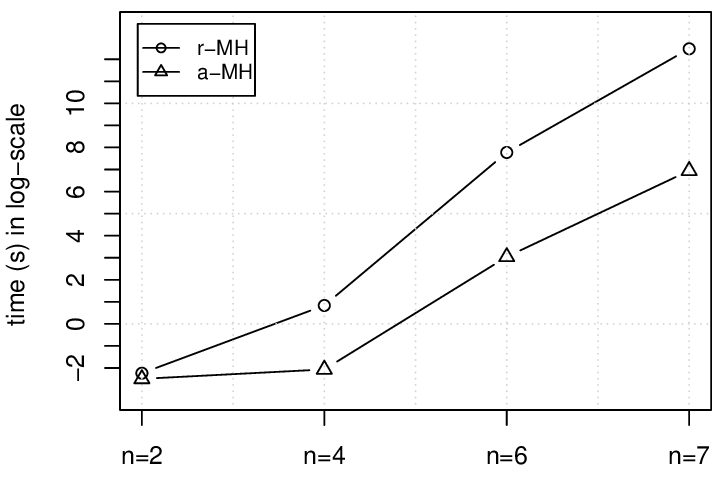}
\caption{Plot to compare the running times (s) in log-scale for 10 steps of two algorithms in the setup of Setting 1, for the qubits $ n =2,4,6,7 $ ($ d= 4, 16, 64, 128 $). }
\label{fg_runtime}
\end{figure}

From Figure \ref{fg_runtime}, it is clear to see that our adaptive MH implementation is greatly faster than the previous implementation from \cite{mai2017pseudo} by at least two orders of magnitude as the number of qubits increase. More specifically,  the data are simulated as in Setting 1 for $n=2,4,6,7$ for which the dimensions of the density matrix are $ d= 4, 16, 64, 128 $ and of the empirical frequencies matrices $ [\hat{p}_{\mathbf{a}, \mathbf{s}}] $ are $ 9\times 4, 81 \times 16,  729\times 64,  2187\times 128 $.  We note that this improvement is quite significant for practical quantum tomography where computational time is a precious resource.

\subsubsection*{Tunning parameters via acceptance rate}
The tunning parameters $ \beta_y ,   \beta_z $ are chosen such that the acceptance rate of Algorithm \ref{al_randomMH} is approximating 0.3, which follows the optimum acceptence probability for random-walk Metropolis-Hastings under various assumptions \cite{gelman1997weak}.  For example, as in our experiments, for $ n=2 $ qubits:  $ \beta_y = 0.33 ,   \beta_z = 0.2 $; for $ n=3 $ qubits:  $ \beta_y = 0.03 ,   \beta_z = 0.03 $ and for $ n=4 $ qubits:  $ \beta_y = 0.03 ,   \beta_z = 0.02 $ (all are run with $\alpha =1,  \lambda = m/2$). We note that as the number of qubits $n$ increase,  these tunning parameters tend to be smaller and smaller to asure that the 0.3 acceptance rate is obtained.  

\begin{figure}[!ht]
\centering
\includegraphics[scale=.6]{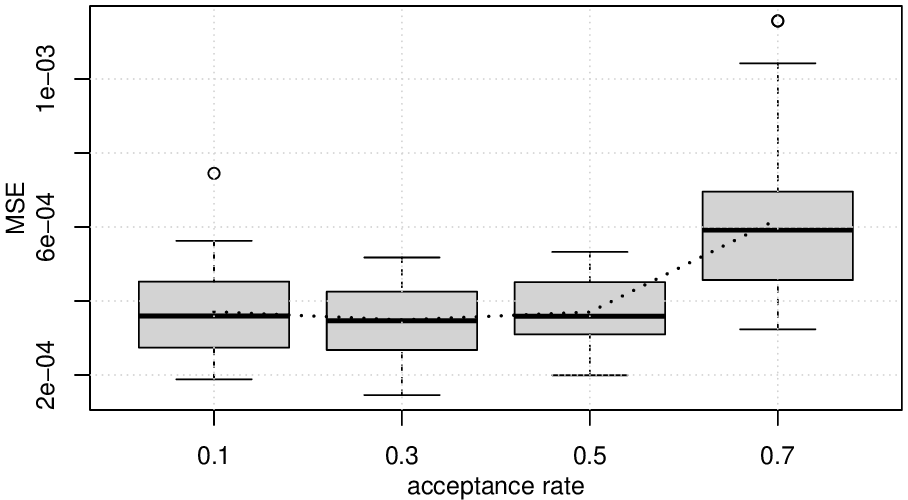}
\caption{Boxplots to examine the effect of the acceptance rate to MSE.  The simulations are run within Setting 2 for $n=4 $.}
\label{fg_ac}
\end{figure}

As an illustration, we conduct some simulations with $n = 4 $ qubits in Setting 2.  It can be seen from Figure \ref{fg_ac} that the acceptance rate around 0.3 would be optimal, where as high acceptance rate like 0.7 could make the algorithm be trapped at local points,  and very small acceptance rate as 0.1 could make the algorithm convergernce slower.

\subsubsection*{Similar accuracy with less variation}

\begin{figure}[!ht]
\centering
\includegraphics[scale=.73]{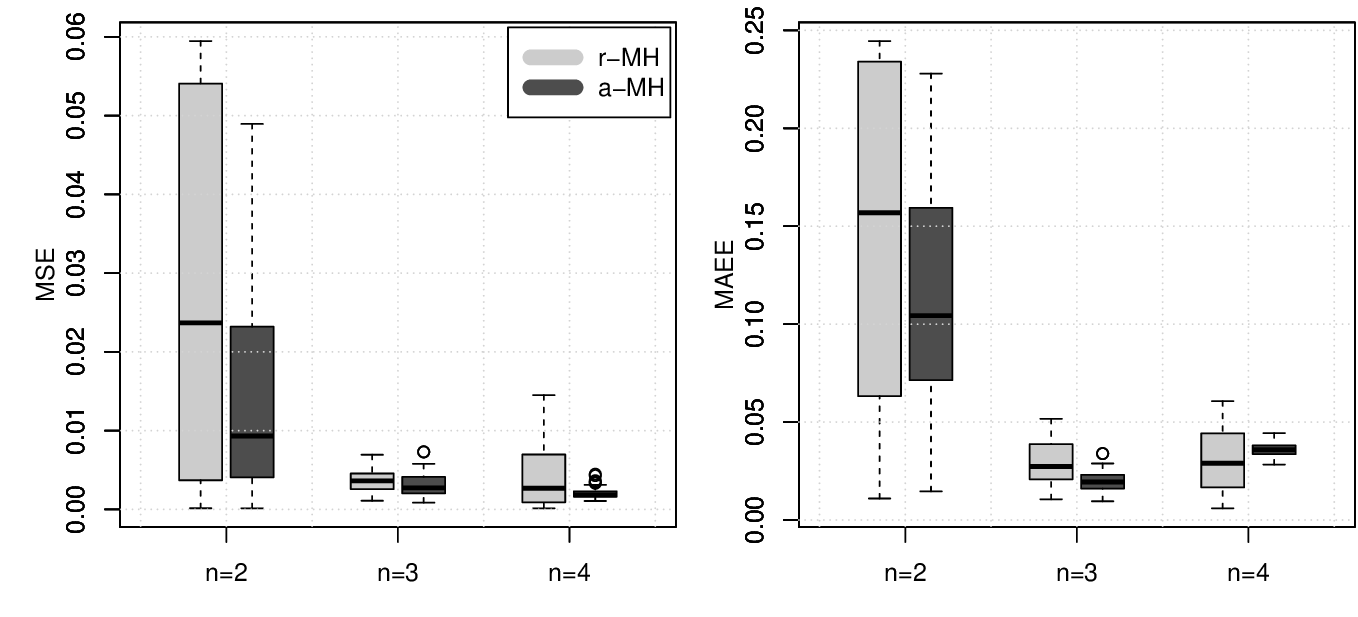}
\caption{MSE}
\label{fg_MSE}
\end{figure}

Results from Figure \ref{fg_MSE} return that both algorithm share similar accuracy in term of both considered errors. However,  it shows a clear improvement that our proposed adaptive algorithm yields much stable results compare to the naive MH approach as expected.

\section{Discussion and conclusion}
\label{sc_conclusion}
We have introduced an efficient sampling algorithm for Pseudo-Bayesian quantum tomography,  especially for the prob-estimator. Our approach is an adaptive Metropolis-Hasting implementation which shows a clear improvement in convergence and computational time comparing with a naive MH implementation.  We would like to mention that such an improvement is significant important for practical quantum state tomography.

Last but not least, faster algorithms based on optimization, such as Variation inference,  for Bayesian quantum tomography would be an interesting research problem. However, it should be noted that the analysis of the uncertainty quantification when using Variational inference is not known, while this matter is an important aspect in the problem of quantum state estimation.

\section*{Availability of data and code}
The R codes and data used in the numerical experiments are available at:  \url{https://github.com/tienmt/bqst} .
\section*{Acknowledgments}
This research of T.T.M was supported by the European Research Council grant no. 742158.



\subsection*{Financial disclosure}

None reported.

\subsection*{Conflict of interest}

The authors declare no potential conflict of interests.




\clearpage
\appendix
\section{Naive Metropolis-Hastings}
\label{appendix}
\begin{algorithm}[!ht]
\caption{MH implementation from \cite{mai2017pseudo}}
For $ t$ from $ 1$ to $ T$, we iteratively update through the following steps:
\begin{description}
\item[updating for $Y_i's $:] for $i $ from $1 $ to $d $,
\\
Sample $\tilde{Y}_i  \sim h(y|Y^{(t-1)}_i) $
where $h$ is a proposal distribution given explicitely below.
\\
Calculate $\tilde{\gamma_i} = \tilde{Y}_i / (\sum_{i=1}^d\tilde{Y}_i) $.
\\
Set
$$ Y^{(t)}_i  =
\begin{cases}
\tilde{Y}_i    &\text{with probability }\min \left\{ 1,R(\tilde{Y}, Y^{(t-1)})\right\},
\\
 Y^{(t-1)}_i        & \text{otherwise} 
\end{cases} $$
where $R(\tilde{Y}, Y^{(t-1)})$ is the acceptance ratio given below.
\\
Put $\gamma_i^{(t)} =Y^{(t)}_i /(\sum_{j=1}^dY^{(t)}_j) , i = 1,\ldots,d$.

\item[updating for $V_i's $:] for $i $ from $1 $ to $d $,
\\
Sample $ \tilde{V}_i $ from the uniform distribution on
 the unit sphere.
\\
Set
$$ V^{(t)}_i  =
\begin{cases}
\tilde{V}_i    &\text{with probability }   \min \{ 1,A(V^{(t-1)}, \tilde{V}) \} ,
\\
 V^{(t-1)}_i        & \text{otherwise}  ,
\end{cases} $$
where $A(V^{(t-1)}, \tilde{V})$ is the acceptance ratio given below.
\end{description}
\end{algorithm}

In details,  $ h(\cdot|Y^{(t-1)}_i)$ is  the probability distribution of $U = Y^{(t-1)}_i \exp(y)$ where $y \sim \mathcal{U}(-0.5,0.5) $.
Following~\cite{robert2013monte} the acceptance ratios are then given by:
\begin{align*}
\log(R( \tilde{Y}, Y^{(t-1)}))
&= \lambda \ell\left(\sum_{i=1}^d \tilde{\gamma_i}
 V_i V_i^{\dagger} ,\mathcal{D}\right)
- \lambda \ell\left(\sum_{i=1}^d \gamma^{(t-1)}_i
V_i V_i^{\dagger} ,\mathcal{D}\right)
\\
& + \sum_{i=1}^d((\alpha-1)\log(\tilde{Y}_i) -\tilde{Y}_i )
-
\sum_{i=1}^d((\alpha-1)\log(Y_i^{(t-1)}) -Y_i^{(t-1)})
\\
& + \sum_{i=1}^d
\tilde{Y}_i - \sum_{i=1}^d Y_i^{(t-1)}
\end{align*}
where $\ell(\cdot,\mathcal{D})$ stands for $\ell^{prob}(\cdot,\mathcal{D})$.

\clearpage
\bibliographystyle{apalike}

\end{document}